# Innovative Weight Simulation in Virtual Reality Cube Games: A Pseudo-Haptic Approach

Woan Ning Lim[1], Edric Leong Yi Junn[1], Yunli Lee[1], and Kian Meng Yap[2]
[1] School of Engineering and Technology, Sunway University, Selangor, Malaysia
[2] Research Centre For Human-Machine Collaboration (HUMAC), School of Engineering and Technology, Sunway University, Selangor, Malaysia
(Email: woanningl@sunway.edu.my)

**Abstract ---** This paper presents an innovative pseudo-haptic model for weight simulation in virtual reality (VR) environments. By integrating visual feedback with voluntary exerted force through a passive haptic glove, the model creates haptic illusions of weight perception. Two VR cube games were developed to evaluate the model's effectiveness. The first game assesses participants' ability to discriminate relative weights, while the second evaluates their capability to estimate absolute weights. Twelve participants, aged 18 to 59, tested the games. Results suggest that the pseudo-haptic model is effective for relative weight discrimination tasks and holds potential for various VR applications. Further research with a larger participant group and more complex scenarios is recommended to refine and validate the model.

## 1 INTRODUCTION

Virtual reality (VR) excels in generating a profound sense of presence and immersion. With advancements in VR and haptic technology, there is increasing interest in simulating virtual object properties such as shape, texture, temperature, stiffness, and weight. However, accurately replicating the sense of weight in a virtual environment remains a significant challenge due to the absence of real gravitational forces. The human process of weight determination is complex, involving multiple sensory systems (touch, visual, and force senses) and encompassing sensory, perceptual, and decisional subprocesses [1].

Haptic interfaces can create a realistic sense of weight by rendering tactile and proprioceptive sensations, yet they are constrained by physical limitations in balancing throughput with the size and weight of actuators. Pseudo-haptic techniques, which exploit human cognitive characteristics, can mitigate these limitations by enhancing weight perception. Nonetheless, reliance solely on visual effects can diminish the realism of virtual object manipulation. The deficiencies in current weight perception simulation techniques indicate a need for a robust, multi-gesture weight perception model adaptable to diverse applications [2].

This paper presents two VR cube games featuring weight simulation using an innovative pseudo-haptic model with a passive haptic glove. The model leverages human cognitive characteristics to induce haptic illusions in weight perception by integrating visual feedback with voluntarily exerted force. Tested through these VR cube games, the model aims to enhance weight perception performance, improve adaptability, and minimize complications associated with active haptic approaches.

## 2 RELATED WORK

The sensory mechanisms underlying weight perception date back to Ernst Weber's early psychological experiments, which emphasized the role of force in perceiving heaviness [3]. Weight discrimination is more effective when an object is actively lifted rather than passively placed on a hand. The perception of an object's weight is theoretically determined by its mass; however, research indicates that weight discrimination is not solely a somatosensory task [4]. Visual information, including size, color, and material, also plays a crucial role in identifying an object's weight, with larger or brighter objects perceived as heavier.

Pseudo-haptic techniques [5], [6], which primarily use visual cues to simulate haptic properties, can create the illusion of weight without active haptic devices. Passive force feedback, exerted and controlled by users with passive isometric devices [7], also contributes to haptic sensation. For example, when a user's visual object movement is artificially slowed, the user exerts more force, creating a sense of resistance.

Lécuyer et al. [8] tested the perception of spring stiffness using an isometric Spaceball and concluded that passive force feedback combined with visual effects can create stiffness sensations. Achibet et al. [9] used grip force to vary perceived effort in tasks like grabbing objects and rotating a dial with a hand exerciser device, demonstrating that users could control virtual objects by manipulating the device. Keller et al. [10] associated virtual object movement with pressing force on the tabletop, although their non-VR environment study lacked quantification of force and weight perception association.

Ponto et al. [11] and Chen et al. [12] found that exertion force scales linearly with object mass, using EMG signals to determine object movement. Their study, conducted in a CAVE environment, showed that muscle activity trends for varying weights were not significant. Hummel et al. [13] proposed two methods using a finger tracking device with a pinch intensity sensor to relate passive force with weight perception. They found that pinching force provided a better realism experience than finger distance due to tactile sensation, though skin type and moisture affected sensor accuracy.

In summary, the integration of passive force feedback and visual cues can enhance the perception of weight in virtual environments, despite challenges in achieving realistic kinesthetic or cutaneous sensations. Further research is needed to refine these methods and improve their application in VR.

## 3  WEIGHT SIMULATION MODEL

An innovative pseudo-haptic model incorporating passive hand force and visual feedback to simulate the perception of weight for virtual objects within a virtual reality (VR) environment was designed and developed. The proposed model consists of two fundamental components: the hand force detection module and the visual feedback module. Unlike other pseudo-haptic research that relies on visual effect based on a static C/D ratio [5], our proposed model [14] uses the passive force feedback to formulate a dynamic C/D ratio rendering to create more complex and continuous visual effects during object manipulation, enhancing both realism and effectiveness.

### 3.1  Hand Force Detection Module

In the hand force detection module, a sensor-enabled wearable device was developed to capture the force applied on the palm and fingers and transfer the signals into the VR environment for visual feedback simulation. The system uses force-sensitive resistors (FSRs) to detect the exerted hand force. An Arduino Uno, or any compatible development platform, acts as an analog-to-digital converter to read the variable resistance of the FSRs. The Arduino IDE is employed for writing code and uploading software to the Arduino board. The Unity Development Platform, along with relevant assets and libraries, serves as the integrated development environment for creating the VR application. The Oculus Quest 2 VR headset is used by the user for testing and experiments. The hand force detection prototype is illustrated in Figure 1.

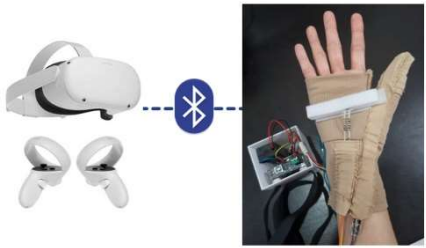

Fig.1  Hand force detection prototype

To create a versatile and realistic hand force detection device capable of accommodating different manipulation styles in object lifting, two specific gestures—pinching and gripping—were targeted. The pinching gesture is more effective for distinguishing lighter objects, while the gripping gesture is better suited for heavier objects. The interface was designed to detect two types of pressures: the pressure on the thumb and the pressure on the palm. To preserve tactile sensitivity and allow freedom of hand movement, all fingers except the thumb remain free from any sensor devices. To optimize sensor performance, a load concentrator (puck) made of compliant and rigid material was used to ensure that 100% of the force load is concentrated within the sensing area.

The FSR signals are transmitted from the Arduino board to the system responsible for visual rendering via Bluetooth using the HC-05 module. The hand force detection circuit diagram is shown in Figure 2. When the Arduino converts the analog voltage (0-5V) to digital, it produces a number between 0 and 1023. The sensors need calibration to correlate their electrical output to force units. In this study, we used the method from Duarte Forero et al. [15] to derive a conversion formula for calculating the forces of pinching and gripping gestures.

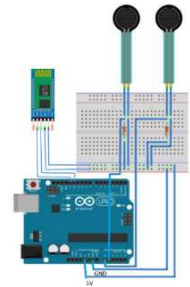

Fig.2  Circuit diagram of the hand force detection module connecting HC-05 module and FSRs to Arduino Board

### 3.2  Visual Feedback Module

The pseudo-haptic model enhances sensory perception by using the brain's processing of visual and auditory information to create the illusion of haptic sensations. Visual cues are crucial in pseudo-haptic design. Previous research [12], [13] used simple binary visual effects (e.g., picked-up or dropped) for weight perception, but this lacked the ability to convey nuanced force changes, resulting in a less immersive experience. Research shows that pseudo-haptic visual effects using the C/D ratio provide the strongest cues and highest weight discrimination accuracy compared to other visual effects like angle of inclination and motion profile [6]. Our approach merges the C/D ratio with passive force feedback to produce more sophisticated and dynamic visual effects in object manipulation.

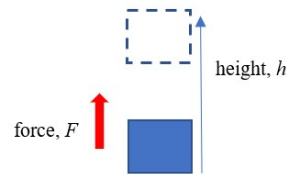

Fig.3  Object lifting to the height of $h$ with force $F$

Newton's second law states that an object's acceleration is directly proportional to the net force and inversely proportional to its mass. To lift an object, an equivalent force to its weight must be exerted, with increased force accelerating it as illustrated in Figure 3. Assuming the hand moves upward in a straight path while lifting, the work done ($W$) is a function of the change in height ($h$) and the exerted force ($F$), as described by the equation (1).

$$W = F \times h \qquad (1)$$

In the visual feedback module, the work done is evaluated by comparing the virtual hand and the real hand tracked by the sensor. The expected work done ($W_{expected}$) is defined as the work of moving the object by the expected height ($h_{expected}$) with the expected force ($F_{expected}$). During lifting, the actual work done ($W_{actual}$) is determined by the exerted force ($F_{actual}$) and the physical height ($h_{actual}$). By matching the actual work done to the expected work done, we find that the expected height ($h_{expected}$) is the displacement of the actual height ($h_{actual}$) by the C/D ratio (R), which is the ratio of the exerted force ($F_{actual}$) to the expected force ($F_{expected}$), as expressed in the equation (2).

$$R = \frac{F_{actual}}{F_{expected}} \qquad (2)$$

In the visual simulation, the object's position is proportional to the ratio of the user's exerted force to the lifting force threshold. This prompts users to adjust their force behavior based on the object's perceived weight. The rules are shown in Table 1.

Table 1  Visual simulation rules

| State | Condition | C/D ratio | Simulation |
|---|---|---|---|
| Resting | Force < weight | < 1 | Stay static |
| Resting | Force >= weight | >=1 | Lift |
| Lifting | Force = weight | 1 | Display position is the same as the physical position. |
| Lifting | Force > weight | > 1 | Display position is higher than the physical position, object accelerates. |
| Lifting | Force < weight | < 1 | Display position is lower than the physical position, object drops. |

## 4  GAME DESIGNS

Two VR cube games were developed to evaluate the effectiveness of the proposed pseudo-haptic model. Twelve participants, aged 18 to 59, were recruited for this experiment, with half playing the games with the C/D ratio on and the other half with it off.

The experimental procedure began with a warm welcome, and participants were informed they could abort or pause the experiment at any time. They then signed a consent form, acknowledging their understanding and agreement to participate. To familiarize participants with the tasks, a demonstration video was shown, introducing the force sensor glove and providing guidance on grasping a cube in VR using both gripping and pinching gestures.

Before starting the VR experiment, participants were equipped with the force sensor glove and a VR headset. Participants were asked to engage in two games: "Arrange the Cubes" and "Balance the Scale." The rules for each game were clearly explained. Participants were allowed unlimited attempts to complete each game and were encouraged to use both pinching and gripping gestures to better perceive the weight.

### 4.1  First Game: Arrange the Cubes

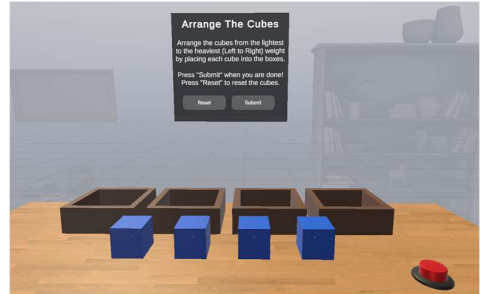

Fig.4   "Arrange the Cubes" game

The first game aims to assess the efficiency of the proposed model in enabling participants to discriminate the relative weight of objects by comparing cubes with different weights. Participants are required to arrange the cubes from lightest to heaviest, placing them in boxes from left (lightest) to right (heaviest). The game setup includes a table with a red button, four blue-colored cubes, four brown containers, and an instruction screen, as illustrated in Figure 4. The cubes have reference weights of 100g, 700g, 1800g, and 2200g, and their positions are randomized.

Once all cubes are placed, participants press the "Submit" button to check their arrangement. If the arrangement is incorrect, an "Incorrect Screen" is shown, and participants can press the "Reset" button to retry, returning the cubes to their original positions. If the cubes are arranged correctly, a "Success Screen" is displayed. The game can be restarted by pressing the "Restart" button, which will randomize the positions of the cubes. If participants decide to give up, they can press the red button on the table.

### 4.2  Second Game: Balance the Scale

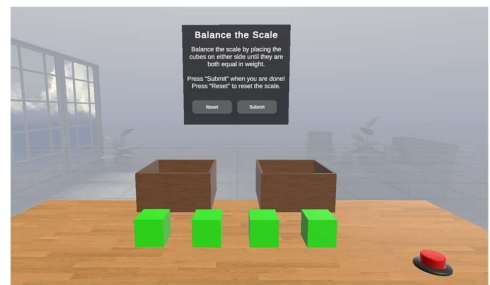

Fig.5   "Balance the Scale" game

The second game evaluates whether the proposed model allows participants to accurately estimate the absolute weight of cubes and correctly place them on a balance scale so that the total weight on both sides of the scale is balanced. The game setup includes a table with a red button, four green-colored cubes, a balance scale with two brown containers (one on each side), and an instruction screen, as illustrated in Figure 5. Unlike the first game where each cube has a different weight, this game features one very light cube weighing 100g, one very heavy cube weighing 2100g, and two medium-weight cubes each weighing 1100g. The starting positions of the cubes are randomized.

During the game, the balance scale remains static until the "Submit" button is pressed. Pressing "Submit" checks if the cubes' total weight is balanced on both sides. If not, an "Incorrect Screen" appears, and participants can press "Reset" to retry, returning the cubes and scale to their original positions. If balanced, a "Success Screen" is displayed. The game can be restarted by pressing "Restart," which resets and randomizes the cubes' positions. Participants can press the red button if they choose to give up.

## 5 CONCLUSION

All participants successfully completed both games. In the first game, 35 attempts were made: 22 with the C/D ratio off and 13 with it on. This indicates that fewer attempts were needed with the C/D ratio, supporting that visual simulation of the C/D ratio with exerted force enhances relative weight discrimination of virtual objects. In contrast, the second game recorded 25 attempts: 9 with the C/D ratio off and 16 with it on. This suggests that the C/D ratio does not facilitate easier estimation of the absolute weight of a cube. The results align with Rietzler et al.[5], indicating that the C/D ratio is more effective for relative weight discrimination than for precise weight estimation. Further analysis of the first game showed that the two lighter cubes were never placed in Containers 3 or 4. The lightest cube was mostly placed in Container 2, and the second lightest in Container 1, while the heaviest cubes were correctly placed. This indicates participants had difficulty distinguishing lighter cubes compared to heavier ones.

While these findings provide valuable insights, the small sample size limits their generalizability. Testing with a larger, more diverse participant group would offer a more robust evaluation of the proposed pseudo-haptic model. Further psychophysical experiments are recommended to quantitatively assess the model's effectiveness. Additionally, the games were limited to four cubes to ensure an appropriate level of difficulty and time constraints. Further testing with increasing number of cubes could provide more comprehensive results.

Overall, this cubes game experiment demonstrates that the proposed pseudo-haptic model successfully generates weight simulation and perception in virtual reality environments. The model's ability to enhance relative weight discrimination suggests its potential for various VR applications.